\documentclass{webofc_modif}
\usepackage[varg]{txfonts}   
\usepackage{amsfonts,amssymb,amsmath,amscd,mathtools,slashed,mathrsfs, physics}
\usepackage{enumerate}

\usepackage[backend=bibtex,citestyle=ieee,style=ieee]{biblatex}
\bibliography{causality_bib}
\usepackage{hyperref}

\wocname{}
\woctitle{ }
\newcommand{\set}[1]{\left\{ {#1} \right\}}

\newcommand{\scal}[1]{\left< {#1} \right>}

\newcommand{\setR}{{\mathbb R}}
\newcommand{\setC}{{\mathbb C}}

\newcommand{\equi}{\Leftrightarrow}
\newcommand{\lequi}{\ \ \Longleftrightarrow\ \ }

\newcommand{\A}{\mathcal{A}}

\newcommand{\C}{\mathcal{C}}

\renewcommand{\H}{\mathcal{H}}

\newcommand{\J}{\mathcal{J}}
\newcommand{\M}{\mathcal{M}}

\newcommand{\T}{\mathcal{T}}

\renewcommand{\bar}{\overline}		

\begin{document}
\selectlanguage{english}
\title{Physical models from noncommutative causality}
%
%

\author{Nicolas Franco\inst{1}\fnsep\inst{2}\fnsep\thanks{\email{nicolas.franco@unamur.be}}}

\institute{Namur Center for Complex Systems (naXys) \& Department of Mathematics, University of Namur,\\ rue de Bruxelles 61, 5000 Namur, Belgium 
\and
Copernicus Center for Interdisciplinary Studies, ul. Szczepa\'nska 1/5, 31-011 Krak\'ow, Poland}

\abstract{
We introduced few years ago a new notion of causality for noncommutative spacetimes directly related to the Dirac operator and the concept of Lorentzian spectral triple. In this paper, we review in a non-technical way the noncommutative causal structure of many toy models as almost commutative spacetimes and the Moyal-Weyl spacetime. We show that those models present some unexpected physical interpretations as a geometrical explanation of the Zitterbewegung trembling motion of a fermion as well as some geometrical constraints on translations and energy jumps of wave packets on the Moyal spacetime.
}
\vspace*{-0.2cm}
\maketitle
\section{Introduction}
\label{intro}

We call noncommutative causality the introduction of a causal order on noncommutative spacetimes. Such an order should exactly correspond to the traditional notion of causality when noncommutativity is removed. The establishment of a well-defined notion of causality on noncommutative spacetimes is clearly not obvious. As an example, the traditional tools of noncommutative quantum field theory fail to define a valid causal structure (without violation) on the Moyal spacetime \cite{MoyalAcausality1,MoyalAcausality2,MoyalAcausality3}. However, the theory of noncommutative geometry based on spectral triples and initially developed by Connes \cite{Connes:1994aa, MC08} allows us to define a valid notion of causality which can be applied on several physical models. Such a notion of causality requires an extension of the theory to spaces with Lorentzian signatures, called Lorentzian noncommutative geometry, which we review in the next section. Then, we present some results obtained on three different physical models: the first two are almost commutative spacetimes and correspond to some partial models coming from the noncommutative Standard Model of particle physics \cite{Dungen} and the last one is the well-known Moyal-Weyl spacetime using the noncommutative $\star$-product and canonical commutation relations.

\section{Lorentzian Spectral Triple and noncommutative causality}
\label{lost}

Traditional tools for (Riemannian) noncommutative geometry are spectral triples $(\A,\H,D)$. In Lorentzian noncommutative geometry, they are replaced by Lorentzian spectral triples. The exact set of axioms of Lorentzian spectral triples is still an active research subject, but the current proposals have many similarities \cite{Pas,Rennie12,F5,BESNARD2017}. We present here the version proposed in \cite{CC2014} since it is the most suitable for guaranteeing the existence of a causal structure.

A Lorentzian spectral triple is given by the data $(\A,\widetilde{\A},\H,D,\J)$
with:
\begin{itemize}
\item A~Hilbert space $\H$,
\item A~non-unital dense $*$-subalgebra $\A$ of a $C^*$-algebra, with a~faithful representation  on
$\H$,
\item A~preferred unitization $\widetilde{\A}$ of $\A$, which is also a dense $*$-subalgebra of a $C^*$-algebra,
with a~faithful representation on $\H$ and such that $\A$ is an ideal of
$\widetilde{\A}$,
\item An unbounded operator $D$, densely defined on $\H$, such that:
\begin{itemize}
\item $\forall a\in\widetilde{\A}$,   $[D,a]$ extends to a~bounded operator on $\H$, \item $\forall
a\in\A$,   $a(1 + \scal{D}^2)^{-\frac 12}$ is compact, with $\scal{D}^2 = \frac 12 (D D^* + D^*
D)$,
\end{itemize} 
\item A~bounded operator $\J$ on $\H$ with $\J^2=1$, $\J^*=\J$, $[\J,a]=0$,
$\forall a\in\widetilde{\A}$ and such that:
\begin{itemize}
\item $D^*=-\J D \J$ on $\text{Dom}(D) = \text{Dom}(D^*) \subset \H$,
\item there exist a densely defined self-adjoint operator $\mathcal{T}$ with $\text{Dom}(\mathcal{T}) \cap \text{Dom}(D)$ dense in $\H$ and with $\left(1+ \mathcal{T}^2 \right)^{-\frac{1}{2}}\in
\widetilde{\A}$, and a positive element $N\in\widetilde\A$ such that $\J  = -N [D,\mathcal{T}]$.
\end{itemize}
\end{itemize}
We say that a Lorentzian spectral triple is even if there exists a~$\mathbb Z_2$-grading $\gamma$ of $\H$ such that
$\gamma^*=\gamma$, $\gamma^2=1$, $[\gamma,a] = 0$ $\forall a\in\widetilde{\A}$, $\gamma
\J =- \J \gamma$ and $\gamma D =- D \gamma $.\\

The operator $\J$ is called the fundamental symmetry. Its role is to turn the positive definite inner product of the Hilbert space $\scal{\cdot, \cdot}$ into an indefinite inner product $(\cdot,\cdot) = \scal{\cdot,\J \cdot}$ (Krein space \cite{Bog}). In fact, the natural inner product for the spinors on a Lorentzian spin manifold is the indefinite one, and the Hilbert space can be constructed using the fundamental symmetry $\scal{\cdot, \cdot} =  (\cdot,\J\cdot)$. The traditional condition on $D$ to be selfadjoint is recovered here by requesting $D$ to be skew-selfadjoint within the Krein space \mbox{$ (D\cdot,\cdot) = (\cdot,-D\cdot)$} which is equivalent to request $D^*=-\J D \J$ on $\H$ \cite{Stro}. However, for a too general operator $\J$, the signature can correspond to a pseudo-Riemannian one \cite{Rennie12}. Hence we impose here the additional condition $\J=-N[D,\mathcal{T}]$ which guarantees that the signature is Lorentzian \cite{CC2014,F5}.

Let us see what happens to this definition when the algebra is commutative. If we consider a locally compact complete\footnote{By complete we understand the following: Let us consider the split metric of the globally hyperbolic spacetime \mbox{$g=-N^2 d^2\T + g_\T$}, then $\M$ is complete under the Riemannian metric $g^R=N^2 d^2\T + g_\T$.} globally hyperbolic spacetime $\M$ of dimension $n$ with a~spin structure $S$, then we can always construct a~commutative Lorentzian spectral triple in the following way:
\begin{itemize}
\item $\H_\M = L^2(\M,S)$ is the Hilbert space of square integrable sections of the spinor bundle over~$\M$ (using the positive definite inner product on the spinor bundle),
\item $D_\M = -i(\hat c \circ \nabla^S) = -i e^{\;\;\mu}_a\gamma^{a} \nabla^S_\mu$ is the Dirac operator\footnote{Conventions
used in the paper are $(-,+,+,+,\cdots)$ for the signature of the metric and
$\{\gamma^a,\gamma^b\}=2\eta^{ab}$ for the flat gamma matrices, with $\gamma^0$ anti-Hermitian and
$\gamma^a$ Hermitian for $a>0$. $e^{\;\;\mu}_a$ stand for vierbeins.} associated
with the spin connection~$\nabla^S$,
\item \label{AM} $\A_\M \subset C^\infty_0(\mathcal{M})$ and $\widetilde{\A}_{\M} \subset
C^\infty_b(\mathcal{M})$ with pointwise multiplication are some appropriate sub-algebras of the algebra of
smooth functions vanishing at infinity and the algebra of smooth bounded functions respectively,
with $\widetilde{\A}_\M$ being be such that $\forall\, a\in\widetilde{\A}$, $[D,a]$ extends to
a~bounded operator on $\H$,
\item $\J_\M=i\gamma^0$, where $\gamma^0$ is the first flat gamma matrix.
\end{itemize}
If $n$ is even, then $\gamma$ is given by the chirality operator.\\

Our goal is to characterize the natural causal order between the points of $\M$. By causal order, we mean that two points are causally related, with $p \preceq q$, if and only if there is a future directed causal curve from~$p$ to~$q$, i.e.~a curve whose tangent vector is causal (almost)-everywhere. This idea is purely geometric and needs to be algebraized using the elements of a Lorentzian spectral triple. The solution comes from the notion of smooth causal functions, which are the smooth real-valued functions non-decreasing along every future directed causal curve. Since $\M$ is globally hyperbolic, the causal structure is completely determined by the cone of smooth causal functions $\C$ in the following way \cite{Bes09}: 
\begin{equation}
\ \forall p,q\in\M,\  p \preceq q   \equi  \forall f\in\C,\ f(p) \leq f(q).
\end{equation}
 The causal functions $f$ have a necessary and sufficient characterization in term of the Dirac operator and the fundamental symmetry with the condition \cite{CQG2013} (see also \cite{Francodist} for a shorter proof): 
 \begin{equation}
 \forall \phi \in \H, \scal{\phi,\J[D,f] \phi} \leq 0.
 \end{equation}

Since noncommutative spacetimes are non-local, hence pointless, we need an alternative definition of the notion of point. Thanks to Gelfand's theory, we can use the concept of pure states on the algebra $\A$ (and their extension to the unitization $\widetilde\A$) as our points on noncommutative spacetimes since their are in one-to-one correspondence with the points of the manifold in the commutative case using the evaluation map $\xi \sim p$ if and only if $\forall f\in\widetilde\A,\ \xi(f)=f(p)$. Hence we can use the following definition of a causal structure for an arbitrary Lorentzian spectral triple:
\begin{equation}
\forall\chi,\xi\in P(\widetilde\A), \ \chi \preceq \xi  \lequi  \forall a\in \C,\ \chi(a) \leq \xi(a),
\end{equation}
 \begin{equation}
 \text{with }\C = \set{ a\in\widetilde\A \ | \ a=a^*,  \forall \phi \in \H \scal{\phi,\J[D,a] \phi} \leq 0 } .
\end{equation}

When the Lorentzian spectral triple is commutative and constructed from a globally hyperbolic spacetime (or even a causally simple spacetime \cite{Ming}), the relation $\preceq$ corresponds to the usual causal relation on $\M$ (the unitization process adds some extra points located at infinity which can just be ignored). This definition can be naturally extended to non-pure states on the algebra, which can represent non-local events or wave packets \cite{Eckstein2017, PhysRevA.95.032106}.

\section{First model: The almost commutative spacetime $\M \times M_2(\setC)$}
\label{mod1}

Almost commutative spacetimes are a tool to naturally integrate gauge theories to usual spacetimes using noncommutative geometry. It is a kind of Kaluza-Klein product between a continuous space and a discrete space. The continuous space $(\A_\M,\widetilde{\A}_\M,\H_\M,D_\M,\J_\M)$ is a commutative even Lorentzian spectral triple representing the spacetime $\M$ while the finite space $(\mathcal{A}_F,\H_F,D_F)$ is built from a noncommutative discrete algebra $\mathcal{A}_F$. The product space is defined as:
\begin{equation}
\A=\A_{\M}\otimes \A_F,\quad \widetilde{\A}={\widetilde{\A}}_{\M}\otimes\A_F,\quad H=H_{\M}\otimes H_F,\quad D=D_{\M}\otimes1+\gamma_{\M}\otimes D_F,\quad \J = \J_{\M} \otimes I.
\end{equation}
 If the finite algebra is chosen to be $\A_F = M_n(\setC)$, the inner automorphisms of the discrete part of the model behave as the gauge group $U(n)$. From this statement comes the noncommutative Standard Model where the discrete space is chosen in order to reproduce the groups $U(1) \times SU(2) \times SU(3)$ \cite{Dungen}.

For this first model, we define an almost commutative spacetime integrating a $U(2)$ gauge field constructed as $\A_{F} = M_2(\setC)$, $\H_{F} =  \setC^{2}$ and $D_{F} = \text{diag}(d_1,d_2)$ with $d_1,d_2\in\setR$, $d_1\neq d_2$. In order to specify the causal structure on it, we need to understand the space of pure states. Since pure states on $M_2(\setC)$ are elements $\xi\in\setC P^1 \cong S^2$ and pure states on $C^\infty_0(\mathcal{M})$ correspond to the points $p\in\M$, a specific pure state on the product spacetime can be identified by $\omega_{p,\xi}$ with $p \in \M$ and $\xi \in S^2$ so the product spacetimes is point by point equivalent to the cartesian product of $\M$ and a sphere.

\begin{figure}[h]
\centering
\sidecaption
\includegraphics[width=8cm,clip]{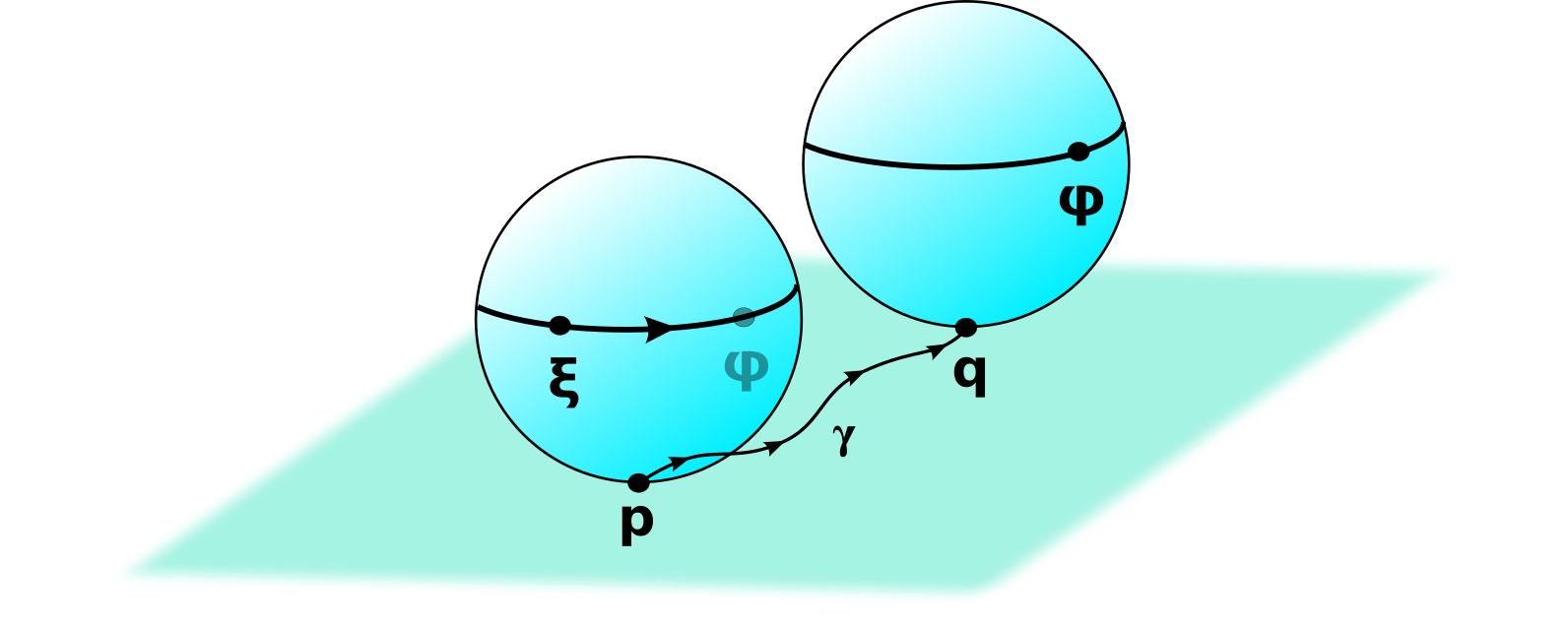}
\caption{Representation of the causal relations on the almost commutative spacetime $\M \times M_2(\setC)$}
\label{fig2spheres}      
\end{figure}

Then we have the following result coming from \cite{SIGMA2014}: Two pure states $\omega_{p,\xi}$ and $\omega_{q,\varphi}$ are causally related, with \mbox{$\omega_{p,\xi} \preceq \omega_{q,\varphi}$}, if and only if:
\begin{itemize}
\item $p \preceq q$ (hence there is no causality violation on $\M$), 
\item They are located on the same parallel of latitude on $S^2$ (hence $\xi$ and $\varphi$ can be described by the variation of they angle $\theta_\varphi-\theta_\xi$ on the given parallel),
\item $\tau(\gamma) \geq  \frac{\abs{\theta_\varphi-\theta_\xi}}{\abs{d_1-d_2}}$
 where $\tau(\gamma)$ represents the proper time of a particle moving along a curve $\gamma : p \rightarrow q$ (cf.~Figure \ref{fig2spheres}).
 \end{itemize}

This first model is a completely toy-model with no direct physical application, except that it can be considered, up to a reduction to the special group, as a part of the noncommutative Standard Model. However, the obtained constraint on the proper time has a physical interpretation as a maximal speed (a kind of speed of light limit) within the internal space. Indeed, since $\abs{\theta_\varphi-\theta_\xi}$ is a distance on the parallel, we have a fixed boundary for the internal speed:
\begin{equation}
\frac{\abs{\theta_\varphi-\theta_\xi}}{\tau(\gamma)} \leq  \abs{d_1-d_2}
\end{equation}
 entirely defined by the eigenvalues of the finite Dirac operator. Such an internal speed of light limit is recurring among models of noncommutative causality.

\section{Second model: The two-sheeted spacetime}
\label{mod2}

The second model is also an almost commutative spacetime but where the discrete algebra $\A_F$ is reduced to the simple non-trivial one: $\A_F = \setC \oplus \setC$. Here the internal space is reduced to two separated points and the product space is just two exact copies of the spacetime $\M$. $\A_F$ is represented on $\H_{F} =  \setC^{2}$ and $D_{F} = \begin{psmallmatrix}0 & m \\m^* & 0\end{psmallmatrix}$ is just defined by a parameter $m\in\setC$ where $\frac{1}{\abs{m}}$ represents the distance between the two points.

\begin{figure}[h]

\centering
\sidecaption
\includegraphics[width=8cm,clip]{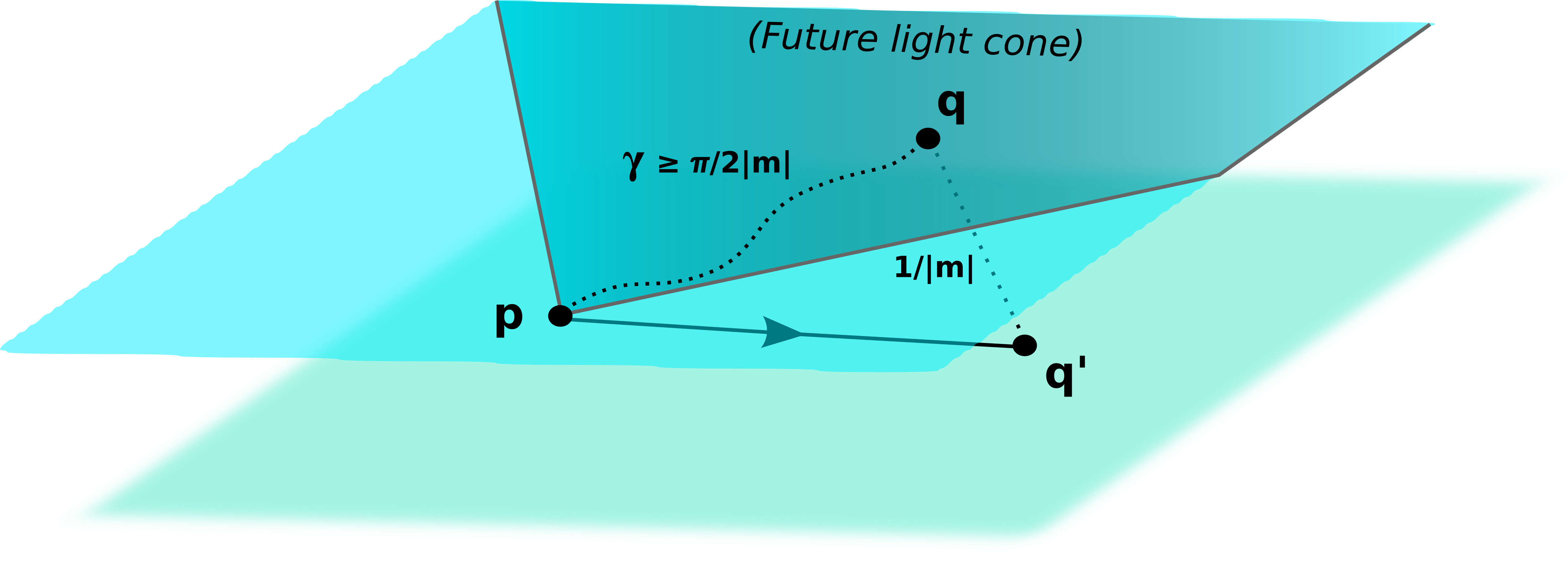}
\caption{Representation of the causal relations on the two-sheeted spacetime $\M \times \setC^2$}
\label{figtwopoints}       
\end{figure}

Then we have a similar result than for the $\M \times M_2(\setC)$ model \cite{JGP2015}: The causal structure is preserved on each sheet and two points $p$ and $q'$ on separated sheets $(+,-)$ are causally related with $p \preceq q'$ if and only if: 
\begin{itemize}
\item They are causally related if considered on the same sheet ($p \preceq q$),
\item $\tau(\gamma) \geq  \frac{\pi}{2\abs{m}}$  where $\tau(\gamma)$ represents the proper time of a particle moving along a curve $\gamma : p \rightarrow q$ (cf.~Figure \ref{figtwopoints}).
\end{itemize}

Once more we recover an internal speed of light limit related to the distance between the two sheets. Of course a pure state does not "jump" from one sheet to another, but goes continuously through a path of mixed-states. This model can easily be extended to a non-constant distance between the sheets with the introduction of a scalar field $\psi$ replacing $m$. Such a scalar field has the behavior of a Higgs field in almost commutative models \cite{MC08,Dungen}. Indeed, the internal speed of light limit is preserved but the proper time is now computed using the Higgs field as a kind of weight:\begin{equation}\label{proper_time}
\tau_\psi(\gamma) = \int_p^q  ds\abs{\psi} \sqrt{-g(\dot \gamma,\dot \gamma)}    \geq\frac \pi 2.
\end{equation}

The question is now: can such a model correspond to some specific physical phenomenon, where a particle is able to switch between two given states $(+,-)$ but with a constraint on the rate of change? As shown in \cite{Zitter}, the answer is given by the "Zitterbewegung". Zitterbewegung, which means "trembling motion", is a rapid oscillation of the value of the velocity and position operator of a fermion obeying the Dirac equation \cite{ZitterSchroedinger}. There exist two interpretations of this movement: an oscillation between positive and negative energy states \cite{Thaller} or an oscillation in chirality \cite{ChiralZitter,ChiralZitter2}. It has been simulated twice \cite{ZitterBEC,ZitterNature,ZitterPhotonics}, with a period of oscillation for a free particle given by:
\begin{equation}\label{TZB}
T_{\text{ZB}} = \frac{\pi \hbar}{m c^2}.
\end{equation}

Our two-sheeted spacetime can be used directly to model the Zitterbewegung behavior, assuming the oscillations occur between the two sheets. We can equip our almost commutative spacetime with the fermionic action: 
\begin{equation}
S_F = \scal{\psi,\J D\psi} = \int_\M {\psi^*_+} D_\M \psi_+  +  {\psi^*_-} D_\M \psi_- + m ( {\psi^*_+} \psi_- + {\psi^*_-} \psi_+) 
\end{equation}
 where we make the distinction between the chirality eigenstates $\psi_{\pm}$. This is the action describing a single Dirac fermion of mass $m$ propagating in a curved spacetime $\M$. If we restore the physical dimensions on the causality constraint, we get:
\begin{equation}\label{constrain_zitter}
(p,+) \preceq (q,-) \lequi p \preceq q\ \text{ and }\ \tau(\gamma) \geq  \frac{\pi\hbar}{2\abs{m}c^2} = \frac{T_{\text{ZB}}}{2}
\end{equation}
where we recover the exact frequency of the Zitterbewegung \eqref{TZB}.

Besides describing in a geometrical way the Zitterbewegung effect on a curved spacetime, the two-sheeted model can go one step ahead. Indeed, we can introduce the inner fluctuations of the Dirac operator  \cite{JGP2015}: 
\begin{equation}
D_A = (D_\M +\gamma^\mu A_\mu) \otimes 1 + \gamma_M \otimes \left(\begin{smallmatrix}0 & \psi\\ \psi^* & 0\end{smallmatrix}\right).
\end{equation}
 The additional vector field $A_\mu$ has absolutely no impact on the causality constraint \eqref{constrain_zitter} while the scalar field $\psi$ introduces the modification of the proper time $\tau_\psi(\gamma)$ defined in \eqref{proper_time}. This predicts an observable impact from a variation of the mass of the particle, which should be of the order $10^{-20}$s, while the model predicts that there should be no impact from an electromagnetic field.\\

However, on can point the fact that the frequency of the Zitterbewegung for a free particle is exactly recovered while taking the maximal speed inside the internal space, as if the particle was forced to follow the maximal speed. This could bring us to rethink about the interpretation of particles in almost commutative models. Indeed, a similar result is obtained in \cite{WatchSake} where a two-sheeted flat spacetime is used in order to study the energy-momentum dispersion relation in a noncommutative geometrical way. It is shown that the boundary of the causal condition $\tau(\gamma) =  \frac{\pi}{2\abs{m}}$ reproduces the relativistic energy-momentum relation for massive and massless fermions, but not when the inequality is relaxed.

So why not making a new postulate that, on almost commutative spacetimes, every particles should "move" at a maximal constant speed $c$, where this speed is calculated from a kind of pythagorean relation between the continuous spacetime and the internal space. If we adopt such a postulate, then massive particles on almost commutative spacetimes can move freely on the continuous spacetime, at a speed with value between zero and strictly less than the light speed, while the "remaining speed" is performed at the level of the internal space, giving the trembling motion of Zitterbewegung as well as the relation $E^2 = (pc)^2 + (mc^2)^2$ where $pc$ corresponds to some energy on the continuous space while $mc^2$ is the energy on the internal space (hence the energy at rest, as a direct consequence of the Higgs field $m=\psi$, with a completely geometrical origin). Also for massless particles, the postulate automatically implies a constant speed $c$ on the continuous spacetime, since the internal space is then reduced to disconnected points and does not allow any movement, as well as the energy relation $E=pc$. However, two models are certainly too few to have any validation of such a postulate, but this could help us in the future to better understand almost commutative spacetimes and make a potential distinction between physical phenomena and mathematical behaviors coming only from the models.

\section{Third model: The Moyal spacetime}
\label{mod3}

The Moyal-Weyl spacetime is a very interesting space on a physical point of view since the noncommutativity given by the star product is total and reproduces the canonical commutation relation $[x^\mu,x^\nu] = i\theta$. Also, the pure states are naturally non-local and we don't need to go to the mixed-states to introduce wave packets. Moreover, Moyal planes fit very well into spectral triples formalism \cite{Gayral}.

For this model we consider a two-dimensional Minkowski spacetime ${\mathbb R}^{1,1}$ with the following noncommutative Lorentzian spectral triple: $\mathcal{H} = L^2({\mathbb R}^{1+1}) \otimes {\mathbb C}^{2}$ with the usual positive definite inner product, $\A$ is the space of Schwartz functions $\mathcal{S}({\mathbb R}^{1,1})$ with the $\star$-product defined as:
\begin{equation}
(f\star g)(x):=\frac{1}{(\pi\theta)^2}\int d^2y\,d^2z\ f(x+y)g(x+z)e^{-2i\,y^\mu\,\Theta^{-1}_{\mu\nu}z^\nu},
\end{equation}
with $\Theta_{\mu\nu}:=\theta\left(\begin{smallmatrix} 0&1 \\ -1& 0 \end{smallmatrix}\right)$, the unitization $\widetilde{\A} = (\mathcal{B},\star) $ is the unital algebra of smooth functions which are bounded together with all derivatives, $D = -i \partial_\mu \otimes \gamma^\mu$  is the usual flat Dirac operator where $\gamma^0 = i\sigma^1$, $\gamma^1 = \sigma^2$ are the flat Dirac matrices and $\mathcal{J}= i\gamma^0$ is the fundamental symmetry.

Functions and states on the Moyal spacetime can be easily described using as orthonormal basis the Wigner eigenfunctions of the two-dimensional harmonic oscillator $a= \sum_{mn} a_{mn} f_{mn}$ where:
\begin{equation}
 f_{mn}={{1}\over{(\theta^{m+n}m!n!)^{1/2}}}{\bar{z}}^{\star m}\star f_{00}\star z^{\star n} \text{ with } \ z=\frac{x_0+ix_1}{\sqrt{2}} \text{ and } f_{00} = 2e^{-\frac{x_0^2+x_1^2}{\theta}}.
 \end{equation}
The pure states are all the normalized vector states on the matrix representation: 
\begin{equation}
\omega_\psi(a)=2\pi\theta\sum_{m,n}\psi^*_ma_{mn}\psi_n ,\quad 2\pi\theta\sum_m|\psi_m|^2=1.
\end{equation}

Since the amount of vector states is huge, we need to restrict to some particular kinds of pure states. In a first time, we only consider the coherent states, which are the vector states defined, for any $\kappa\in{\mathbb C}$, by:
\begin{equation}
\varphi_m = \frac{1}{\sqrt{2\pi\theta}}e^{-\frac{{\left\vert \kappa\right\vert }^2}{2\theta}} \frac{\kappa^m}{\sqrt{m!\theta^m}}.
\end{equation}

The coherent states correspond to the possible translations under the complex scalar $\sqrt{2}\kappa$ of the ground state $\ket{0}$ of the harmonic oscillator, using the correspondence $\kappa\in{\mathbb C} \cong {\mathbb R}^{1,1}$. They are the states that minimize the uncertainty equally distributed in position and momentum. The classical limit of the coherent states, when $\theta \rightarrow 0$, corresponds to the usual pure states on ${\mathbb R}^{1,1}$, hence to the points of the usual Minkowski spacetime \cite{MoyalDist3}.

\begin{figure}[h]
\centering
\sidecaption
\includegraphics[width=8cm,clip]{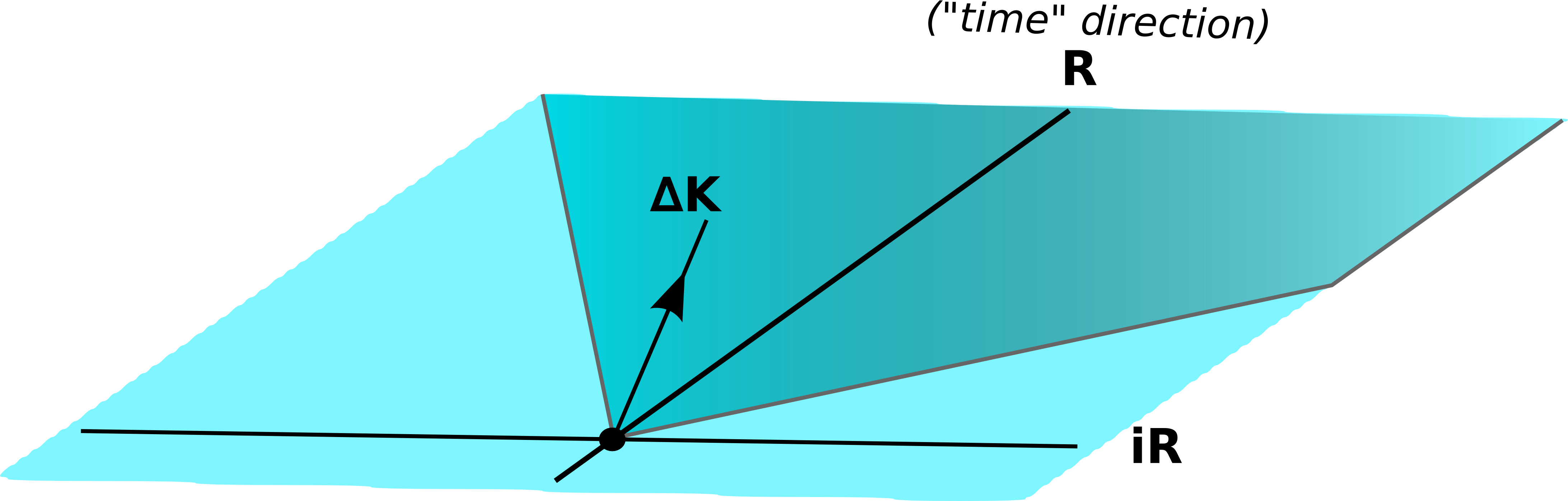}
\caption{Representation of the causal relations by translation between coherent states on the Moyal spacetime}
\label{figmoyalCS}       
\end{figure}

For the coherent states, we find the following causal structure \cite{FrancoMoyal}: Let us suppose that two coherent states $\omega_\xi,\omega_\varphi$ are defined by the two complex scalars $\kappa_1,\kappa_2\in{\mathbb C}$, then those coherent states are causally related, with $\omega_\xi \preceq\omega_\varphi$, if and only if $\Delta\kappa=\kappa_2-\kappa_1$ is inside the convex cone of ${\mathbb C}$ defined by $\lambda=\frac{1+i}{\sqrt2}$ and $\bar\lambda = \frac{1-i}{\sqrt2}$  (i.e.~the argument of $\Delta\kappa$ is within the interval $[-\frac{\pi}{4},\frac{\pi}{4}]$, cf.~Figure \ref{figmoyalCS}).

Hence, with the identification between ${\mathbb C}$ and ${\mathbb R}^{1,1}$, we recover the exact future and past light cones of Minkowski, but with the difference that we do not consider movements of points but translations of Gaussian functions. Hence we are naturally dealing with non-local states.  In such a case, we can define a kind of "time direction" as translations under positive real scalars $\kappa$.\\

However, the ground state $\ket{0}$ is not the only one which can be translated, since every energy level $\ket{n}$ produces a similar causal structure. Those translations of the basic eigenstates of the harmonic oscillator, which we can write $\alpha_{\kappa}\ket{n}$ for a $\kappa$ translation, are called generalized coherent states. Hence, those states behave like a infinite number of copies of the Minkowski spacetime. The study of the complete causal structure of the generalized coherent states, and especially the causal relations between the different energy levels, is a work in progress and will be the subject of a future paper, but we can already mention the following nice result:

If ${\Delta\kappa}\in\setR$ is such that:
\begin{equation}
 {\Delta\kappa} \geq \frac{\pi}{2} \sqrt{\frac\theta 2} \frac 1{\sqrt{n+1}},
 \end{equation}
then $\ket{n} \preceq \alpha_{\Delta\kappa} \ket{n+1}$ and $\ket{n+1} \preceq \alpha_{\Delta\kappa} \ket{n}$.

\begin{figure}[h]
\centering
\sidecaption
\includegraphics[width=8cm,clip]{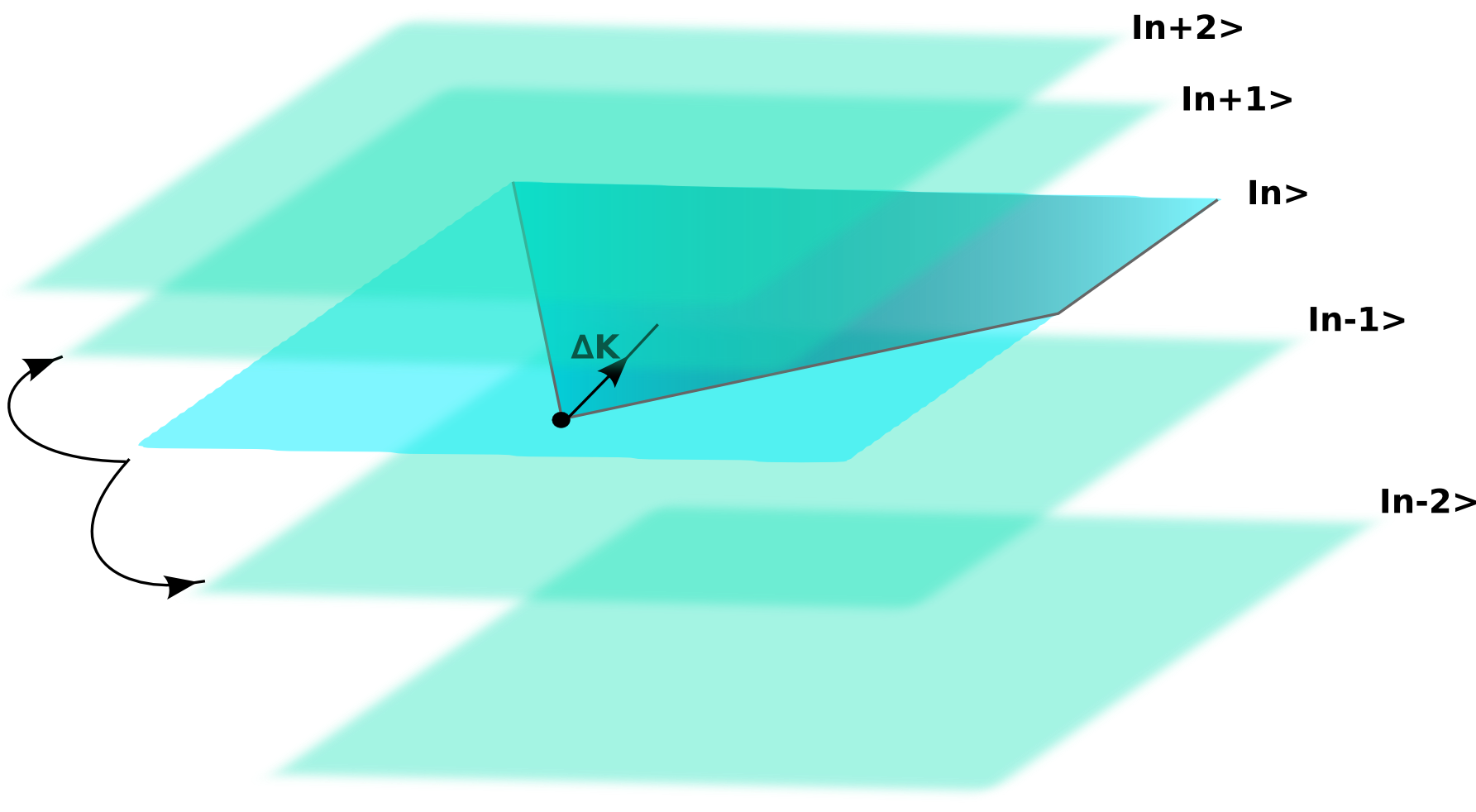}
\caption{Representation of the causal relations between general coherent states from different energy levels on the Moyal spacetime}
\label{figmoyalGCS}       
\end{figure}

Hence we recover a similar result than for the two-sheeted model which is reproduced between each pair of energy level (the causal relations between every sheet is deduced by transitivity) with a distance between consecutive energy levels $\ket{n}$ and $\ket{n+1}$ being equal to $\frac {\sqrt{\theta}}{\sqrt 2\sqrt{n+1}}$ (cf.~Figure \ref{figmoyalGCS}). 

When taking the limit \mbox{$\theta \rightarrow 0$}, we recover the usual causality on Minkowski with all sheets merged and usual translations of points.

This model represents in a completely geometrical way waves packets under causal translations with a lower bound on time in order to change the energy level. However, the testability of this lower bound is out of sight since it should be of the order of the Plank time, which is somehow logical from a quantum point of view.

\begin{acknowledgement}
Publication supported by the John Templeton Foundation Grant "Conceptual Problems in Unification Theories" (No. 60671).
\end{acknowledgement}

\printbibliography


\end{document}